\documentclass[12pt, a4paper]{article}
 
\usepackage[hmargin=2.5cm,vmargin=3cm]{geometry}
\usepackage{graphicx}
\usepackage{amsmath}
\usepackage{enumerate}
\usepackage{graphicx}  
\usepackage{dcolumn}   
\usepackage{bm}        
\usepackage{amssymb}   
\usepackage{enumerate}
\usepackage{amsmath}

\begin{document}
\title{The thermodynamics of a black hole in equilibrium implies the breakdown of Einstein equations on a macroscopic near-horizon shell}
\author {Charis Anastopoulos\footnote{anastop@physics.upatras.gr} and Ntina Savvidou\footnote{ksavvidou@patras.gr}     \\
 {\small Department of Physics, University of Patras, 26500 Greece} }
\maketitle

\abstract{
We study a black hole of mass $M$, enclosed within a spherical box, in equilibrium with its Hawking radiation. We show that the spacetime geometry inside the box is described by the Oppenheimer-Volkoff equations for radiation, except for a {\em thin shell} around the horizon. We use the maximum entropy principle to show that the invariant width of the shell is of order $\sqrt{M}$,  its entropy is of order $M$ and its temperature of order $1/\sqrt{M}$ (in Planck units). Thus, the width of the shell is {\em much larger} than the Planck length.
Our approach is to insist on thermodynamic consistency when classical general relativity coexists with   the Hawking temperature in  the description of a gravitating system. No assumptions about an underlying theory  are made and no restrictions are placed on the origins of the new physics near the horizon. We only employ classical general relativity and the principles of thermodynamics. Our result is strengthened by an analysis of the trace anomaly associated to the geometry inside the box, i.e., the regime where quantum field  effects become significant corresponds to the shells of maximum entropy around the horizon.}

\maketitle
\flushbottom

\section{Introduction}

The attribution of thermodynamic properties to black holes  is incompatible with classical general relativity (GR) \cite{Bek, Hawk}.   The derivation of Hawking radiation requires a quantum treatment of matter degrees of freedom. For this reason, the origin of black hole thermodynamics is commonly sought at the  quantum gravity level.

In this article, we focus on the  thermodynamic level of description for black holes. Thermodynamics is a theory for macroscopic
  coarse-grained variables and it can be consistently formulated without any reference to the underlying physics. In particular, thermodynamics applies even if the underlying physics is fully quantum.
For this reason, we believe that  it is possible to formulate a thermodynamic description of black holes that {\em incorporates the quantum effects of matter} within a classical theory of gravity.  In Ref. \cite{SavAn14}, we showed that
  the thermodynamics of gravitating systems in equilibrium is holographic at the classical level, in the sense that  all thermodynamic properties are fully specified by variables defined on the system's boundary.
In Ref. \cite{AnSav12}, we constructed a consistent thermodynamic description of solutions to Einstein's equations that correspond to radiation in a box.

In this work, we employ these solutions in order to describe
a black hole of mass $M$ inside a box, in thermal equilibrium with its Hawking radiation.
  We find that  the breakdown of classical GR takes place in a thin shell around the horizon. Since the principles of thermodynamics are insensitive to the microscopic underlying dynamics, we identify the shell's physical characteristics by employing the maximum entropy principle. We find that the shell is characterized by high temperature (of order $1/\sqrt{M}$) and its invariant thickness is of order $\sqrt{M}$.  Hence, the width of the shell around the horizon is {\em much larger} than the Planck scale. This is unlike most existing models that postulate a shell or membrane of Planck-length around the horizon---for example \cite{hooft85,STU93}. The  invariant width of the shell derived here is also larger than the
  invariant distance of $M^{1/3}$ from the horizon characterized by strong gravitational interactions
due to the “atmosphere” of high angular-momentum particles, derived in Ref. \cite{CEIMP96}.

  The electromagnetic (EM) field is an excellent example by which to demonstrate our perspective.   The quantum EM  field has a consistent statistical mechanical description, while the classical EM field has none. Nonetheless, the thermodynamics of the classical EM field is well defined: the equation of state  follows from the classical action, and the entropy functional is inferred from the equation of state. The only imprint of   quantum theory is the Stefan-Boltzmann constant that appears as a phenomenological parameter in the entropy functional. In analogy, when seeking an integrated description of black hole thermodynamics and   GR at the macroscopic level, we  expect that   quantum   effects  are incorporated into  phenomenological parameters   of the thermodynamic potentials.

The structure of this article is the following. In Sec. 2, we describe the background for studying a black hole inside a box. In Sec. 3, we present the properties of the solutions to the Oppenheimer-Volkoff equation inside the box. In Sec. 4, we derive the condition for the breakdown of the Oppenheimer-Volkoff equation showing that it is restricted into a thin shell around the horizon. In Sec. 5, we implement the maximum-entropy principle, using minimal modeling assumptions, in order to identify the properties of the shell. In Sec. 6, we discuss the physical origins of the breakdown of Einstein's equations, and its relation to quantum vacuum fluctuations of matter fields. In Sec. 7, we summarize and discuss our results.

\section{The equilibrium black hole.}
 A black hole in an asymptotically flat spacetime is  not an equilibrium system  because it radiates. However, a black hole enclosed within a perfectly reflecting spherical box is an equilibrium system because it involves two competing processes: emission of Hawking quanta, and their re-absorption after reflection from the boundary.  One expects that the equilibrium state corresponds to the black hole coexisting with its Hawking radiation.  This system has been studied before \cite{Dav78, York85}, albeit with   simplifying assumptions.

Since the Hawking emission of massive particles is exponentially suppressed \cite{Hawk}, radiation is well described by the thermodynamic equations for ultra-relativistic particles:
\begin{eqnarray}
\rho = b T^4,  \hspace{1cm} P = \frac{1}{3}\rho, \hspace{1cm} s = \frac{4}{3} b^{1/4} \rho^{3/4},
\end{eqnarray}
where $\rho$ is the energy density, $P$ is the pressure, $T$ is the temperature and $s$ is the entropy density; $b$ is the Stefan-Boltzmann constant that takes the value $\frac{\pi^2}{15}$ for pure EM radiation.  (We use Planck units, $\hbar = c = G = 1$.) Particle numbers    are not preserved  in the processes of black hole formation and evaporation; thus, they do not define thermodynamic variables and the associated chemical potentials vanish.

Assuming spherical symmetry,   the metric outside the box  is a Schwarzschild solution with Arnowitt-Deser-Misner (ADM) mass $M$. An observer outside the box has access to several  macroscopic variables that are constant in absence of external intervention. Such variables are the   mass $M$,  the area $4 \pi R^2$ of the box,
 the boundary temperature $T_R$ and the boundary pressure $P_R$. The internal energy of a spherically symmetric system coincides with the ADM mass $M$ \cite{SavAn14}.  A change  $\delta R$ of the boundary radius corresponds to work $-P_R (4 \pi R^2) \delta R/\sqrt{1 - 2M/R}$ as measured by a local static observer, or $-P_R (4 \pi R^2)$ to an observer at infinity. The first law of thermodynamics then becomes

\begin{eqnarray}
\delta M = T_{\infty} \delta S - P_R (4 \pi R^2) \delta R
\end{eqnarray}
where $T_{\infty} = T_R/\sqrt{1 - 2M/R}$ is the temperature  at infinity.
The first law above implies that the thermodynamic state space of the system  consists of the variables $M$ and $R$.

This physical system is characterized by two phases, the radiation phase and the black-hole phase. For fixed $R$, and for sufficiently small values of $M$, the box contains only radiation; for higher values of $M$ the box contains a black hole coexisting with
  its Hawking radiation. A heuristic description of the two phases is the following.
  For $2M/R << 1$,  gravity is negligible in the radiation phase, the system is homogeneous with constant density $\rho = m /(\frac{4}{3} \pi R^3)$ and the  entropy is
  \begin{eqnarray}
  S_{rad} = \frac{4}{3} \pi R^3 s = \frac{4}{3} \left(\frac{4}{3} \pi b\right)^{1/4} M^{3/4} R^{3/4}.
  \end{eqnarray}
   For $2M/R$ approaching unity, almost all the mass is contained in the black hole of radius $ 2M$, hence, the Bekenstein-Hawking formula for the black hole entropy applies, $S_{BH} = 4 \pi M^2$. The black hole phase is entropically favored if $S_{BH} > S_{rad}$, i.e., for
   \begin{eqnarray}
   M^5 R^{-3} > \frac{4 b}{3^5 \pi^3}.
\end{eqnarray}

 The radiation phase was studied in Ref. \cite{AnSav12}. In what follows, we construct the  thermodynamics of  the black-hole phase through the following steps:  (i) we derive the  geometry inside the box using classical GR; (ii) since radiation cannot coexist in equilibrium with a horizon in GR,
 we identify the spacetime region where Einstein' s equations break down; (iii) we find an effective macroscopic description for the physics of this region by using the maximum-entropy principle.

\section{Classical geometry inside the box.}

The region inside the box where the black hole coexists with its Hawking radiation  corresponds to  a static solution to Einstein's equations with radiation,

\begin{eqnarray}
ds^2 = - (1 - \frac{2M}{R}) \sqrt{\frac{\rho(R)}{\rho(r)}} dt^2 + \frac{dr^2}{1 - \frac{2m(r)}{r}} + r^2 d\Omega^2, \label{ds2}
\end{eqnarray}
where $ d \Omega^2 = (d \theta^2 + \sin^2\theta d \phi^2)$ and
  $(t, r, \theta, \phi) $ are the standard coordinates. The mass function $m(r)$  satisfies $
\frac{dm}{dr} = 4 \pi r^2 \rho$, and
the energy density $\rho(r)$ satisfies the Oppenheimer-Volkoff (OV) equation
\begin{eqnarray}
\frac{d \rho}{dr} = -\frac{4 \rho}{r^2} \frac{(m+ \frac{4}{3} \pi r^3 \rho)}{1 - \frac{2m}{r}}. \label{OV}
\end{eqnarray}
  We change the variables to
  \begin{eqnarray}
\xi := \ln \frac{r}{R}, \\
u := \frac{2m(r)}{r}, \label{udef} \\
v := 4 \pi r^2 \rho, \label{vdef}
\end{eqnarray}
 to obtain
\begin{eqnarray}
\frac{du}{d\xi} = 2v - u \label{equ} \hspace{1cm}
\frac{dv}{d\xi} = \frac{2v (1 - 2 u - \frac{2}{3}v)}{1-u}. \label{equv}
\end{eqnarray}

 Eq. (\ref{equv})  is to be integrated from the boundary ($\xi = 0$, or $r = R$) inwards, because  the  thermodynamic variables $M$ and $R$ are defined at the boundary. We denote the values of $u$ and $v$ at the boundary as $u_R$ and $v_R$, respectively. Thus, $u_R = 2M/R$ and $v_R = 4 \pi b R^2  T_R^4$.

There are two classes of solutions  to Eq. (\ref{equv}) that are distinguished by their behavior as $r \rightarrow 0$ \cite{box, AnSav12}. The first class contains  solutions with a conical singularity at the center. They satisfy $\rho(0) = 0$ and $m(0) = -M_0$, for some   constant $M_0 > 0$. The solutions in the second class  are regular (everywhere locally Minkowskian). They satisfy $m(0) = 0 $ and $\rho(0) = \rho_c$, for some constant $\rho_c >0$.

The integration of Eq. (\ref{equv}) from the boundary inwards does not encounter a horizon ($u = 1$), except for the trivial case of $v_R = 0$ that corresponds   to a Schwarzschild horizon and no radiation inside the box \cite{AnSav12}. However, there is a  sub-class of singular solutions with  $u \simeq 1$ near a surface $r = r_* $.
These solutions arise for $v_R << u_R$, i.e., for  low density at the boundary.
  We   call these geometries Approximate-Horizon (AH) solutions.

 Next, we study the properties of the AH solutions.
Plots of $u$ and $v$ as a function of $r$  are given in Fig. 1.
A typical AH solution is characterized by three regions
\begin{enumerate}[(i)]

 \item In  region $I$, $u$ increases  and $v$ decreases with decreasing $r$.   $P$ is the local minimum of  $v$.

 \item
In   region $II$, $u$ keeps increasing with decreasing $r$ until it reaches a maximum {\em very close to unity} at $O_*$ ($r = r_* \simeq 2M$); $v$ also increases with decreasing $r$ in region $II$ and equals $\frac{1}{2}$ at $O_*$. By Eq. (\ref{vdef}),   the density
\begin{eqnarray}
\rho_*\simeq \frac{1}{32 \pi M^2}, \label{ro*}
\end{eqnarray}
 and the local temperature
\begin{eqnarray}
T_* = \frac{1}{(32\pi b)^{1/4} \sqrt{ M}} \label{t*}
 \end{eqnarray}
 at $O_*$ depend  only on $M$, within an excellent approximation.

\item Region $III$ corresponds to decreasing $u$; $v$ increases dramatically shortly after   $O_*$, but then drops to zero at $r = 0$.
\end{enumerate}

\begin{figure}[tbp]
\includegraphics[height=9cm]{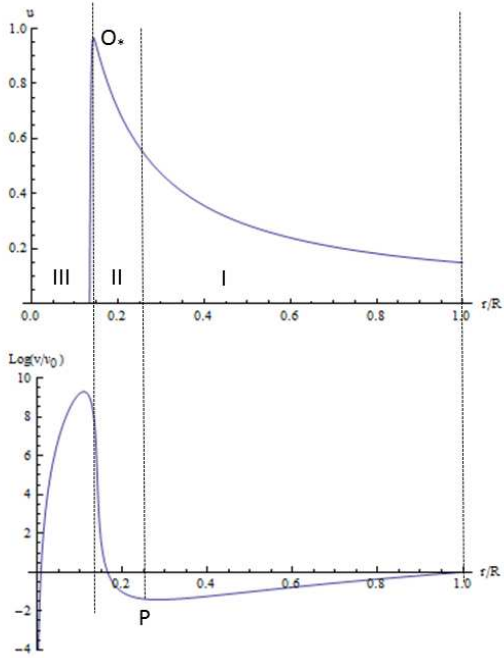} \caption{ \small $u$ and $v$ as functions of $r$ for an AH solution with $u_R = 0.15$ and  $v_R = 0.01$.    Note that we have to use a logarithmic scale for $v$ due to its rapid increase near $O_*$.}
\end{figure}

 An approximate evaluation of the AH solutions is described in Appendix A. Every $AH$ solution is characterized  by the parameter
 \begin{eqnarray}
 \epsilon_* = 1 - u(r_*) << 1 \label{estar0}
 \end{eqnarray}
  that defines the maximal blue-shift at $O_*$. In Appendix A, we express  $\epsilon_*$ as a function of the boundary variables,
  \begin{eqnarray}
  \epsilon_* = \frac{16}{9} u_R (1 - u_R) \sqrt{2 v_R}. \label{estar0a}
  \end{eqnarray}
 Solving Eq. (\ref{estar0a}) for $v_R$ and using Eq. (\ref{vdef}), we relate the boundary temperature $T_R$ to $\epsilon_*$,

 \begin{eqnarray}
T_{R} = \frac{3 \sqrt{\epsilon_*}}{4 \sqrt{2M} (8 \pi b)^{1/4}\sqrt{1 - u_R}}. \label{bt0}
\end{eqnarray}


In the Appendix A, we also prove a relation between radial coordinate at $O_*$ and $\epsilon_*$,
\begin{eqnarray}
r_* = 2M \left( 1 + \frac{3 \epsilon_*}{8} \right),
\end{eqnarray}
and a relation between the value of the mass function at $O_*$, $m_* := m(r_*)$ and $\epsilon_*$,
\begin{eqnarray}
m_* =
 M \left( 1 - \frac{5 \epsilon_*}{8} \right).
\end{eqnarray}

In the vicinity of $O_*$, the metric Eq. (\ref{ds2}) becomes
\begin{eqnarray}
ds^2 = - N_*^2  dt^2 + \frac{dx^2}{\epsilon_* + \frac{x^2}{3M^2 \epsilon_*}}
+ r_*^2 d \Omega^2 \label{ds2b}
\end{eqnarray}
where $x = r - r_*$, and $N_* = \frac{3}{4}\sqrt{\epsilon_*}$ is the lapse function. Interestingly, the proper acceleration at $O_*$ equals $(4M)^{-1}$, i.e., it coincides with the surface gravity of a Schwarzschild black hole of mass $M$.

\section{Breakdown of the OV equation.}
The regions $I$ and $II$ of an AH solution describe the geometry of
  the black hole phase at some distance from the horizon. Since  the OV equation cannot account for the presence of a horizon, it must break down somewhere
 in region $II$, and close to $O_*$. It must be substituted by a different equation that is compatible with the formation of an horizon. However, any such modification must be very drastic: the OV equation is compatible with a horizon only for matter configurations with negative pressure \cite{ST}.

It is conceivable that the equation of state for  radiation   is modified near $O_*$ in order to incorporate quantum effects of non-gravitational origin, such as QED vacuum polarization.
 However, such modifications are unlikely to lead to the negative pressures that are necessary for the formation of a horizon.
For a solar mass black hole, $\rho_* \sim 10^{16}\rho_{H_2O}$, where $\rho_{H_2O}$ is the density of water. Hence,
 $\rho_*$ is of the same order of magnitude with the density at the center of a neutron star. The corresponding local temperature $T_*$ is of the order of  $10^{12}{}^oK$, which is a typical temperature for quark-gluon plasma.  No existing model of strong/nuclear interactions suggests the possibility of
   negative pressure in these regimes. For  super-massive black holes, $\rho_* \sim 10^2 \rho_{H_2O}$; negative pressures are even more implausible in this regime. For this reason, we expect that quantum effects at high densities
      may cause quantitative changes in the thermodynamics of self-gravitating radiation, but they are not strong enough
    to generate a black hole phase. In further support of this assertion, we   note that any contribution from quantum effects would have strong and complex dependence on the mass $M$, involving masses and thresholds from high energy physics. The resulting thermodynamics would not manifest the simplicity and universality of the Bekenstein-Hawking entropy.

Since high densities or temperature cannot lead to the formation of a horizon, the main
  cause  for
  the breakdown of the OV equation in region $II$ is the extreme blue-shift $\epsilon_*^{-1/2}$.
  At extreme blue-shifts, the
     description of matter in terms of hydrodynamic variables (e.g, energy density) fails because the hydrodynamic description is not fundamentally continuous but presupposes a degree of  coarse-graining.

     In Minkowski spacetime, the energy density $\rho$ is defined as $\rho = U/L^3$, where $U$ is the energy in a cube of    size $L$. $L$ defines the degree of spatial coarse-graining and it cannot be  arbitrarily small \footnote{ For thermal radiation at temperature $T$, the requirement that the energy fluctuations are much smaller than the mean energy in a volume $L^3$ implies that $L T >> 1$. At higher temperatures, the Compton wave-length of the electron defines an absolute lower limit to $L$. }.
            The energy density can be treated as a continuous field only  when
     measured at scales much larger than $L$. The hydrodynamic description fails when the fluid dynamics generate length-scales of order $L$. Then, either the consideration of fluctuations or a microscopic treatment is necessary.

 In  curved spacetimes, the coarse-graining   scale $L$ is defined with respect to the local rest frame, so it represents a proper length.
By Eq. (\ref{ds2b}),  the coarse-graining scale  $L$  corresponds to a radial distance $\Delta r \sim  L \sqrt{\epsilon_*}$   near $O_*$. Hence, if $|r_* - 2 m_*| \sim L \sqrt{\epsilon_*}$, or, equivalently, if
\begin{eqnarray}
 M \sqrt{\epsilon_*} \sim L \label{bdown}
 \end{eqnarray}
the hydrodynamic fluctuations obscure any distinction of   $O_*$ from a genuine horizon. We note that Eq. (\ref{bdown}) does not require $L$ to be a constant. A temperature dependence of $L$ is equivalent to a  dependence on the mass $M$, because the local temperature at $O_*$ depends only on $M$. Then, Eq. (\ref{bdown}) still provides an estimate of $\epsilon_*$ as a function of $M$.

An alternative justification of Eq. (\ref{bdown}) is the following. In a hydrodynamic system, local densities and temperature are meaningfully defined only if they vary at scales significantly larger   than the coarse-graining scale $L$; the variation within a shell of volume $L^3$ must be a small fraction of the averaged value.
Tolman's law implies that the product of the local temperature $T$ and the lapse function $N$ is constant. Using   Eq. (\ref{lapse}) for the lapse at $O_*$,
\begin{eqnarray}
\left|\frac{\nabla_rT}{T}\right| = \frac{2}{3M \epsilon_*}.
\end{eqnarray}
The coordinate distance $\Delta r$ corresponding to proper length $L$ near $O_*$ is $\Delta r = L \sqrt{\epsilon_*}$.

When the variation of temperature in a cell of proper length $L$ is of the same order of magnitude as the temperature, the hydrodynamic description breaks down.     The relevant condition is  $|\nabla_r T/T| \Delta r \sim 1$, which implies Eq. (\ref{bdown}).

In Sec. 6, we will show that in the regime that corresponds to the thermodynamically stable black hole, contributions to the stress-energy tensor from QFT in curved spacetime (the trace anomaly) become important. The present analysis is compatible with this result, because in this regime quantum fluctuations of the stress-energy tensor are very strong, and thus, no classical hydrodynamic variables can be defined---see, Sec. 6.3.


\section{Maximum-entropy principle.}
The fundamental thermodynamic variables of the system are the ADM mass $M$ and the box radius $R$. However, the solutions to Einstein equations depend on {\em three} independent parameters, which can be chosen as the mass $M$, the box radius $R$, and the blue-shift parameter $\epsilon_*$. By Eq. (\ref{bt0}), the dependence on $\epsilon_*$ is equivalent to a dependence on the   boundary temperature $T_R$. The equilibrium configuration is determined by the {\em maximum-entropy principle}: the value assumed by any unconstrained parameter in a thermodynamic system is the one that maximizes the entropy subject to the system's constraints \cite{Callen}.

The thermodynamic constraints for an {\em isolated} box is the constancy of $M$ and $R$; the blue-shift parameter $\epsilon_*$ is unconstrained. Hence,
the equilibrium configuration for fixed $M$ and $R$ corresponds to the value of $\epsilon_*$ that maximizes the entropy functional. We expect that the entropy functional  has  one local maximum for each phase.

 The radiation phase maximum has  the larger value of $\epsilon_*$. For $ \sqrt{\epsilon_*} >> L/M$, the OV equation holds everywhere and we recover the thermodynamics of self-gravitating radiation   \cite{AnSav12}.
 Smaller values of   $\epsilon_*$ correspond to the black hole phase. For
 $\sqrt{\epsilon_*} \sim L/M$,  the OV equation breaks down near the surface $O_*$. This breakdown is accompanied by a formation of a horizon $H$ near $O_*$, at $r = r_H < r_*$.
 The violation of the OH equations is restricted to a thin shell around    $O_*$ with a radial width $\delta r := r_* - r_H$ of order  $\epsilon_* M$.  All properties of the shell  depend on   $\epsilon_*$, and they are fully specified once $\epsilon_*$ is fixed by the maximum-entropy principle.

 We   model the spacetime geometry of the black-hole phase as follows. In the region between the bounding box and the surface $O_*$, the metric is described by an AH solution. A horizon is formed at $r = r_H < r_*$   and a thin   shell where the OV equation does not apply extends from $r_H$ to $r_*$.
We write
\begin{eqnarray}
r_H = 2M( 1 - \lambda \epsilon_*),
\end{eqnarray}
 where $\lambda > \frac{5}{8}$ is an unspecified constant of  order  unity.
  The simplification involved in this model is that we assume the breakdown of the OV equation to occur sharply at $O_*$, rather than considering a gradual degradation. This approximation should not affect the order-of-magnitude estimate of the shell's  properties. Note that we need not assume that the shell extends up to the horizon $r = r_H$. This would be problematic because points of the horizon are causally disconnected from the interior. For the subsequent analysis,  it suffices that the shell extends up to a distance from the horizon that is much smaller than $r_* - r_H$.

The total entropy within the box is a sum of three terms,

\begin{eqnarray}
S_{tot} = S_H + S_{sh} + S_{rad},
\end{eqnarray}
where

\begin{enumerate}[(i)]

 \item $S_H$ is the Bekenstein-Hawking entropy of the horizon:
 \begin{eqnarray}
 S_H = \pi r^2_H \simeq  4 \pi M^2 - 8 \pi \lambda \epsilon_* M^2.
 \end{eqnarray}

\item
  The entropy $S_{sh}$ of the thin shell is
 expected to depend only on the local temperature at $O_*$ (hence, on $M$) and on the shell width $\delta r$. For $\delta r = 0$, there is no shell, so $S_{sh} = 0$. It follows that
 \begin{eqnarray}
 S_{sh}(M, \delta r) = \frac{\partial S_{sh}}{\partial \delta r}(M, 0) \delta r + O[(\delta r)^2],
 \end{eqnarray}
 so we write
  \begin{eqnarray}
    S_{sh} \simeq \epsilon_* M B,
    \end{eqnarray}
     where   $B$ is a function of $M$ to  be determined later.

  \item The entropy of radiation $S_{rad}$  is the volume integral of the entropy density $s$ in the regions I and II,
      \begin{eqnarray}
S_{rad} = \frac{4}{3} (4 \pi b)^{1/4} \int_{r_*}^R \frac{r^{1/2} v^{3/4}}{\sqrt{1-u}} dr. \label{srad}
\end{eqnarray}

In the Appendix B, we show that
 \begin{eqnarray}
S_{rad} = \frac{1}{12} (8 \pi b)^{1/4} (2M)^{3/2} \sqrt{\epsilon_*} [1 + O(\epsilon_*)]. \label{srad2}
\end{eqnarray}

\end{enumerate}

Hence, in the regime of small $\epsilon_*$, the total entropy is
\begin{eqnarray}
S_{tot} =  4 \pi M^2 + \frac{1}{12} (8 \pi b)^{1/4} (2M)^{3/2} \sqrt{\epsilon_*} - (8 \pi \lambda M^2 - B) \epsilon_* + O(\epsilon_*^{3/2}), \label{stot2}
\end{eqnarray}
i.e., it is approximated by a polynomial of second order with respect to $\sqrt{\epsilon_*}$.

In an isolated box, the values of $M$ and $R$ are constrained, while $\epsilon_*$ may fluctuate. Hence, the equilibrium configuration is defined as the maximum of the total entropy $S_{tot}$  with respect to $\epsilon_*$.
The maximum occurs for
\begin{eqnarray}
\sqrt{\epsilon_*} = (8 \pi b)^{1/4} \frac{\sqrt{2M}}{12(8 \pi \lambda  M - B) }.
 \end{eqnarray}

 By Eq. (\ref{bt0}), the corresponding boundary temperature is

\begin{eqnarray}
T_R = \frac{1}{16(8 \pi \lambda M - B)\sqrt{1-u_R}}. \label{t0a}
\end{eqnarray}

The boundary temperature should coincide with the Hawking temperature   $T_{\infty} = \frac{1}{8 \pi M}$, blue-shifted by a factor $\sqrt{1-u_R}$. It is a non-trivial check of our model that the $R$ dependence of Eq. (\ref{t0a}) is compatible with such an identification for
 $ B = (8 \lambda - \frac{1}{2}) \pi M$. Then,  the entropy functional, Eq. (\ref{stot2}) is expressed solely in terms of known parameters,

\begin{eqnarray}
S_{tot}(M, R, \epsilon_*) = 4 \pi M^2
+ \frac{(2 \pi b)^{\frac{1}{4}} M^{\frac{3}{2}}}{3}  \sqrt{\epsilon_*} -  \frac{\pi M^2}{2}   \epsilon_* ,  \label{stot}
\end{eqnarray}
and the equilibrium configuration corresponds to
\begin{eqnarray}
\sqrt{\epsilon_*} = \frac{(2 \pi b)^{1/4} }{3 \pi \sqrt{M} }. \label{estar}
\end{eqnarray}

  Eq. (\ref{estar}) implies that $N_* T_* = T_{\infty}$, i.e., Tolman's law is satisfied  for the  Hawking temperature at infinity. This agrees with the results of Refs. \cite{SavAn14, KM75}, where Tolman's law is derived solely from the maximum-entropy principle and it is independent of the dynamics of GR.

The equilibrium configuration Eq.
 (\ref{estar}) must also satisfy the condition
$L \gtrsim \sqrt{\epsilon_*} M $ for the existence of a black hole phase. By Eq. (\ref{estar}),
$ L \gtrsim \sqrt{M}$, i.e., the coarse-graining scale $L$ defines an upper bound to the mass of a black hole that can be nucleated in a box.
This bound is not particularly  restrictive:  it is satisfied  even  by super-massive black holes for $L$  in the atomic scale.

The width $\delta r$ of the shell in the equilibrium configuration is
\begin{eqnarray}
\delta r   =  \left(\frac{3}{8} + \lambda \right) \frac{2\sqrt{2\pi b}  }{9 \pi^2},
\end{eqnarray}
i.e., it is of the order of the Planck length.  However, the proper width $l$ of the shell is by no means Planckian. Eq. (\ref{ds2b}) implies that $l \sim \delta r/ \sqrt{\epsilon_*} \sim \sqrt{M}$.


The   entropy of the shell in the equilibrium configuration is

\begin{eqnarray}
S_{sh} = (8 \lambda  - \frac{1}{2}) \frac{\sqrt{2\pi b}}{9 \pi} M.
\end{eqnarray}
We  estimate the internal energy $E$ of the shell by treating the shell as a single thermodynamic object of temperature $T_{sh} = (\partial S_{sh}/\partial E)^{-1}$.  In thermal equilibrium, $T_{sh}$ should coincide with the local temperature $T_*$ of radiation, Eq. (\ref{t*}). Hence, we obtain
\begin{eqnarray}
 E = \frac{(8\lambda - \frac{1}{2}) (2\pi b)^{1/4}}{9 \pi} \sqrt{M},
\end{eqnarray}
i.e., the internal energy of the shell is proportional to $\sqrt{M}$
modulo a constant of order unity.

\section{Physical origins of the shell}
The results of the previous section follow solely from thermodynamic arguments.  Here, we examine the physical origins of the breakdown of Einstein's equations near $O_*$. First, we examine the classical geometry of the shell by interpolating between the approximate and the true horizon $H$. Then, we examine the possibility that the breakdown of the geometry is due to quantum vacuum fluctuations. We show that the conformal anomaly becomes comparable to the classical stress energy tensor near $O_*$, for $\epsilon_* \sim M^{-1}$, a result that is non-trivially compatible with the condition (\ref{estar}) that follows from the maximum entropy principle. We also argue that the thermodynamic approach presented here can, in principle, resolve existing  problems in the consistent formulation of the quantum back-reaction to the black hole  geometry.

\subsection{The classical geometry of the shell}
The thermodynamic analysis of Sec. 5 estimates the proper length of the shell to be of order $\sqrt{M}$, and hence, much larger than the Planck length. This implies that, in spite of the breakdown of Einstein's equations near $O_*$, a description of the shell in terms of a classical geometry is still possible.

For this reason, we consider
  a spherically symmetric metric, with a mass function that interpolates between the horizon $H$ at $r = r_H$ and the approximate horizon  $O_*$ at $r = r_*$. We assume a power-law interpolation,
\begin{eqnarray}
m(r) = \frac{1}{2}r_H + k (r - r_H)^{a+1}, \mbox{for   }   r_H < r < r_*, \label{interp}
\end{eqnarray}
where
  $k$ and $a$ are positive constants.

  We require that $m(r)$, Eq. (\ref{interp}) is joined with an AH solution at $O_*$, such that the metric and its first derivatives are continuous. This implies
  that $m(r_*) = m_*$, Eq. (\ref{m*}) and that $m'(r_*) = \frac{1}{2}$.  The horizon is defined by  the condition  $2m(r_H) = r_H$. We further require that
   $m'(r_H) = 0$. This means that the effective `density' on the horizon vanishes, because otherwise any matter on the horizon would be causally disconnected from other matter. The last condition implies that $a >0$.

   Then, we obtain

\begin{eqnarray}
k = \frac{1}{2 (a + 1) (2M (1 + a^{-1}) \epsilon_*)^a} \\
r_H = 2M (1 - \frac{5}{8}\epsilon_* - \frac{1}{a} \epsilon_*).
\end{eqnarray}
Hence, the width of the shell is $\delta r = r_* - r_H =  2 M \epsilon_* a^{-1}$.

The proper length $l$ of the shell is
\begin{eqnarray}
l = \int_{r_H}^{r_*} \frac{dr}{\sqrt{1 - \frac{2m(r)}{r}}} = \int_0^{\delta r} dx \frac{\sqrt{r_H+x}}{\sqrt{x}\sqrt{1 - kx^a}} \nonumber \\  \simeq
\sqrt{r_H} \int_0^{\delta r}  \frac{dx}{\sqrt{x}\sqrt{1 - kx^a}} = C(a) \sqrt{2M \delta r},
  \label{properl}
\end{eqnarray}
where
\begin{eqnarray}
C(a) = \int_0^1 \frac{dy}{\sqrt{y} \sqrt{1 - \frac{1}{2(a+1)}\left(\frac{y}{a+1}\right)^a}}
\end{eqnarray}
is a constant of order unity: for example, $C(\frac{1}{2}) = 2.16, C(1) \simeq 2.04, C(2) \simeq 2$.

For $\epsilon_*$ given by Eq. (\ref{estar}), the proper length $l$ is indeed of the order of $\sqrt{M}$.

Comparing Eq. (\ref{interp}) with the OV equation, we can estimate an effective "equation of state" that parameterizes the properties of the shell. The OV equation in the shell is well approximated by
\begin{eqnarray}
\frac{dP}{dz} \simeq - \frac{(\rho + P)}{2z } (1 + 32 \pi M^2 P), \label{OV7}
\end{eqnarray}
where $z = r - r_H$. Numerical solution of Eq. (\ref{OV7}) leads to an effective equation of state, i.e., a relation between $\rho$ and $P$, as shown in Fig. 2. We note that for $a \geq 1$, the effective equation of state is reasonably well approximated by a linear relation of the form  $P = -w \rho$, where $w > 0$.

\begin{figure}[tbp]
\includegraphics[height=5cm]{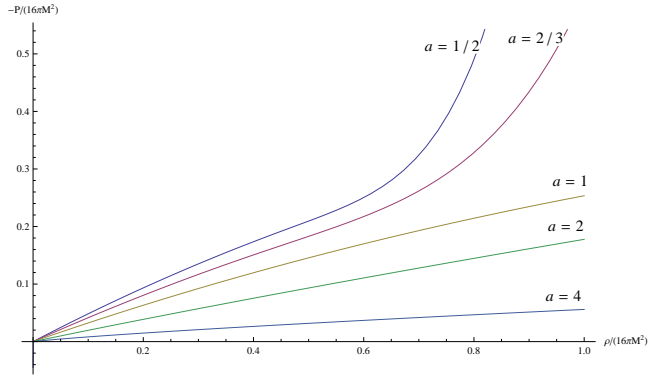} \caption{ \small The effective equation of state inside the shell, the absolute value of the pressure $-P$ as a function of the density $\rho$,  for different values of the interpolation exponent $a$.}
\end{figure}

Near the horizon ($z = 0$),    a linear equation of state with negative pressure is a good approximation for all $a$
\begin{eqnarray}
P =  - \frac{1}{2a+1} \rho, \label{eqst}
\end{eqnarray}

Substituting Eq. (\ref{eqst}) into the continuity equation
\begin{eqnarray}
\frac{dN}{N} = - \frac{dP}{\rho+P},
\end{eqnarray}
we derive the lapse function near the horizon
$N \sim \rho^{1/a} \sim \sqrt{r -r_H}$. $N$ can be expressed as
\begin{eqnarray}
N = \kappa x,
\end{eqnarray}
where $x = \sqrt{\frac{r-r_H}{8M}}$ and $\kappa$ is the surface gravity of the horizon.  Then, the geometry near the horizon

\begin{eqnarray}
ds^2 = - \kappa^2 x^2 dt^2 +  dx^2 + (2M)^2 d\Omega^2, \label{shellm}
\end{eqnarray}
 is of the Rindler type with acceleration $\kappa$.

 We have no analytic expression for $\kappa$, but we expect it to be of order $1/M$. Indeed, if  Eq. (\ref{eqst}) were a good approximation to the effective equation of state throughout the shell, we  would obtain
$\kappa = \frac{3\sqrt{a}}{16 M}$.

\subsection{Relation to the trace  anomaly}

In Sec. 4, we gave a general argument that the usual hydrodynamic notion of the stress-energy tensor fails near the $O_*$,  and, thus, the classical Einstein equations are not reliable near $O_*$. We did not discuss the  physical mechanisms underlying this breakdown. The leading candidate is, of course, quantum phenomena that are expected to be amplified near a high-blue-shift surface.

Since  radiation is   scale invariant at the classical level, the stress energy tensor satisfies $T^{\mu}_{\mu} = 0$. Thus, the size of the quantum effects in  curved spacetime is quantified by the trace anomaly \cite{confan},
\begin{eqnarray}
\Theta =  \langle \hat{T}_{\mu}^{ \mu}\rangle_R,
\end{eqnarray}
i.e., the renormalized expectation value of the trace of a composite operator $ \hat{T}_{\mu \nu}$ that represents the quantum stress-energy tensor.

For a conformally invariant quantum field,
\begin{eqnarray}
\Theta = \alpha {\cal H} + \beta {\cal G} + \gamma \square R, \label{tranomaly}
\end{eqnarray}
where
\begin{eqnarray}
{\cal H} =  R_{\mu \nu \rho \sigma}R^{\mu \nu \rho \sigma} - 2 R_{\mu \nu}R^{\mu \nu} + \frac{1}{3}R^2.  \label{htran}\\
{\cal G} = R_{\mu \nu \rho \sigma}R^{\mu \nu \rho \sigma} - 4 R_{\mu \nu}R^{\mu \nu} +  R^2 \label{gtran}
\end{eqnarray}
are expressed in terms of the Riemann tensor $R_{\mu \nu \rho \sigma}$, the Ricci tensor $R_{\mu \nu}$ and the Ricci scalar $R$; $\square = \nabla_{\mu} \nabla^{\mu}$.

The coefficients $\alpha, \beta$ and $\gamma$ depend upon the spin of the field. Different methods give the same values to all coefficients except $\gamma$. In what follows, we will choose the value of $\gamma$ given by dimensional regularization, $\gamma  = \frac{2}{3} \alpha$. For the EM field, the coefficients $\alpha$ and $\beta$ are
\begin{eqnarray}
\alpha = \frac{1}{160 \pi^2}  \hspace{2cm} \beta = - \frac{31}{2880\pi^2}. \label{alphabeta}
\end{eqnarray}

For a macroscopic black hole, we expect the quantum effects to be significant near the high blue-shift surface $O_*$.  We write the metric as $ds^2 = - f(r)dt^2 + h(r) dr^2 + r^2 d\Omega^2$, and we express the functions $f$ and $h$ as Taylor series in $(r-r_*)$, using Eqs. (\ref{fh*1}---\ref{fh*5}).

The dominant contribution to the curvature tensors near $O_*$ are
\begin{eqnarray}
R_{trtr} &=& - \frac{1}{2} fh R, \hspace{0.3cm}R_{tt} = - \frac{1}{2}fR, \hspace{0.3cm} R_{rr} = \frac{1}{2}hR,
\\
R &=& \frac{f''}{fh} - \frac{(f')^2}{2f^2h} - \frac{f'h'}{2fh^2}.
\end{eqnarray}
All other components of the Riemann tensor are smaller by a factor of $\epsilon_*$.

Substituting into Eqs. (\ref{htran}---\ref{gtran}), we obtain
\begin{eqnarray}
{\cal H} &=& \frac{1}{3}R^2 \\
{\cal G} &=& 0 \\
\square R &=& \frac{1}{h}R'' + \left(\frac{f'}{2fh} - \frac{h'}{2h^2}\right) R'.
\end{eqnarray}

We calculate the Ricci scalar and its derivatives at $O_*$,
\begin{eqnarray}
R_*= 0 \hspace{0.5cm} R_*' = \frac{16}{27} (2M)^{-3} \epsilon_*^{-2}    \hspace{0.5cm} R_*'' = -\frac{64}{27}(2M)^{-4} \epsilon_*^{-3}.
\end{eqnarray}

It follows that in the vicinity of $O_*$, the trace anomaly $\Theta$ is of the order of $(M^4 \epsilon_*^2)^{-1}$. For concreteness, we compute  the value of $\Theta$ at $O_*$,
\begin{eqnarray}
\Theta_* = -\frac{160}{81} \frac{\gamma}{(2M)^4 \epsilon_*^2} = - \frac{320}{243}  \frac{\alpha}{(2M)^4 \epsilon_*^2}.
\end{eqnarray}
By Eq. (\ref{ro*}), we find the ratio
\begin{eqnarray}
\frac{\Theta_*}{\rho_*} = - \frac{640\pi \alpha}{243M^2 \epsilon_*^2}.  \label{thetaro}
\end{eqnarray}
When $\Theta_*/\rho_* $ becomes of order unity,  Einstein's equations fail near $O_*$ due to the quantum effects associated to the trace anomaly.  By Eq. (\ref{thetaro}), the violation of Einstein's equations occurs for
 $\epsilon_* \sim M^{-1}$.  Remarkably, this estimation is in agreement with the value of $\epsilon_*$ determined from the maximum entropy principle, Eq. (\ref{estar}).
 Hence, the  configuration  that maximizes entropy is also characterized by the onset of
violations to Einstein's equations due to the trace anomaly.
 We conclude that  {\em two very different types of argument suggest the same order of magnitude for $\epsilon_*$}, and, implicitly,  the same order of magnitude for the width of the shell.

Substituting the value Eq. (\ref{estar}) into Eqs. (\ref{thetaro}) and (\ref{alphabeta}), we find for pure EM radiation
\begin{eqnarray}
\frac{\Theta_*}{\rho_*} = -\frac{5}{2},
\end{eqnarray}
which suggests that the approximation of a sharply defined shell is only good for order of magnitude estimations. A realistic treatment ought to take into account the gradual deterioration of Einstein's equations as $\epsilon$ becomes smaller.

\subsection{Problems in formulating  back-reaction }

In the previous section, we showed that the trace anomaly is of the correct magnitude to account for the breakdown of the classical Einstein equations near the horizon. The question then arises how to formulate the constitutive equations for the system that incorporate the contribution of the trace anomaly. This is the well-known back-reaction problem for QFT in curved spacetime.

A common proposal for the treatment of back-reaction involves the use of the semi-classical Einstein equations
\begin{eqnarray}
G_{\mu \nu} = 8\pi \left( T_{\mu \nu} + \langle \hat{T}_{\mu \nu}\rangle_R \right). \label{backreaction}
\end{eqnarray}
 In Eq. (\ref{backreaction}), one includes the expectation value of the renormalized quantum stress-energy tensor as source of the gravitational field in addition to a classical distribution of matter. This is clearly an approximation and not  a fundamental theory \cite{PaGl}, since Eq. (\ref{backreaction}) equates a classical observable with a quantum expectation value.

We believe that this approach does not work for the problem at hand, for the following reasons.
\begin{enumerate}
\item The approximation involved in Eq. (\ref{backreaction}) requires that the higher moments of the stress-energy tensor are negligible in comparison to the mean value $\langle \hat{T}_{\mu \nu}\rangle_R $. This is not true, in general. The ratio of energy-density variance to the mean value of the energy density may become of order unity and larger \cite{HuPh97}; in particular, this is the case for the stress-energy fluctuations in Schwarzschild spacetime \cite{HuPh0103}. This behavior is not particular to curved spacetimes , but rather, it is a general feature of the quantum definition of the stress-energy tensor. As such, it persists even in the non-relativistic regime \cite{AnHu15}. One proposed resolution to this problem is to include the quantum fluctuations as a stochastic force in the semiclassical Einstein equations---see, \cite{stochgrav} and references therein.

In fact, the existence of strong quantum fluctuations in the stress-tensor near $O_*$ is compatible with our analysis in Sec. 4 of the breakdown of classical hydrodynamics on $O_*$.  Classical hydrodynamics  presupposes a coarse-grained level of description at which quantum fluctuations are negligible; thus it is incompatible with a regime where quantum fluctuations dominate.

In conventional thermodynamics, the fluctuations of hydrodynamic variables are assumed to be negligible---as long as the observables are averaged within a sufficiently large volume.  Typically, the relative size of fluctuations decreases with $N^{-1/2}$, where $N$ is the  number of particles in the averaging volume. However, this option is not available when considering {\em vacuum} fluctuations encoded in the trace anomaly.
In our opinion, a thermodynamically consistent treatment of such fluctuations requires the definition of a coarse-grained version of the quantum stress-energy tensor operator $\hat{T}^{coarse}_{\mu \nu}$, possibly significantly different from the standard definitions of $\hat{T}_{\mu \nu}$ in the context of QFT in curved spacetime. The key conditions in the definition of $\hat{T}^{coarse}_{\mu \nu}$ are (i)   $\hat{T}^{coarse}_{\mu \nu}$ should be  a quasi-classical variable, i.e., a coarse-grained variable that satisfies  classical evolution equations \cite{GeHa}, and  (ii) the associated   fluctuations should be relatively small so that thermodynamic variables can be properly defined. For examples of quasi-classical hydrodynamic variables defined in  quantum systems, see, Ref. \cite{quasi} and for a discussion of back-reaction in relation to quasi-classical variables, see Ref. \cite{Ana01}.

 \item The trace anomaly, Eq. (\ref{tranomaly}) involves terms up to fourth order  of the metric, while  Einstein equations involve up to second order derivatives. The space of solutions of Eq. (\ref{backreaction}) contains therefore additional variables that correspond to the values of the third and fourth derivative of the metric. However, such variables do not have an obvious physical significance; in particular,  they have no interpretation in terms of known thermodynamic variables. We have no criterion for assigning values to them at the boundary, and thus the solutions to the back-reaction equations are severely under-determined.

 \item Einstein's equations for a static spacetime  correspond to the maximum of the entropy for fixed boundary conditions  \cite{SavAn14}. An {\em ad hoc} modification of  Einstein's  equations (especially one that involves higher derivatives of the metric) is not guaranteed to satisfy this property. This is  problematic, because it implies that the geometry obtained from the solution of Eq. (\ref{backreaction}) may not be stable under microscopic fluctuations.
\end{enumerate}

Nonetheless, it is instructive to compute the renormalized expectation value of the stress-energy tensor $\langle \hat{T}_{\mu \nu}\rangle_R$. Closed expressions for $\langle \hat{T}_{\mu \nu}\rangle_R$ for static spacetimes have been computed in the bibliography \cite{Page82, BrOt86, FrZe87}, as well as expressions particular to static spherically symmetric spacetimes \cite{AHS95}. In what follows, we employ the expression  for a thermal stress-energy tensor by Page \cite{Page82}.   This is obtained from a Gaussian path-integral approximation to the field propagator \cite{BePa81}.

The quantum expectation value of the stress energy tensor consists of two terms. One term contains a logarithm of  the lapse function $N$, the other one does not. The presence of the  logarithmic term implies that $\langle \hat{T}_{\mu \nu}\rangle_R$ is not   invariant under a constant conformal transformation $g_{\mu \nu}\rightarrow c g_{\mu \nu}$ for some constant $c$.  For massless fields, this implies an ambiguity in the definition of $\langle \hat{T}_{\mu \nu}\rangle_R$, equivalent to the introduction of an undetermined renormalization mass $\mu$. It turns out that $\langle \hat{T}_{\mu \nu}\rangle_R$ contains a term proportional to $\log(N\mu)$.

We calculate  $\langle \hat{T}_{\mu \nu}\rangle_R$ at $O_*$ as in \cite{Page82} and obtain
\begin{eqnarray}
\bar{\rho}_* = - \langle \hat{T}_{t}^t\rangle_R = \frac{1}{(2M)^4 \epsilon_*^2} \left[ 8 \alpha [\frac{32}{243} - \frac{32}{81} \log(3 \mu \sqrt{\epsilon_*}/4)] + \beta \frac{96}{81} + \gamma \frac{64}{81} \right] \label{rrr}
\\
\bar{P}^r_* = \langle \hat{T}_{r}^r\rangle_R = \frac{1}{(2M)^4 \epsilon_*^2}  \left[ - 8 \alpha \frac{16}{243} \log(3 \mu \sqrt{\epsilon_*}/4)+ \beta \frac{32}{81} + \gamma \frac{32}{243} \right] \\
\bar{P}^{\theta}_* =  \langle \hat{T}_{\theta}^{\theta}\rangle_R  = \frac{1}{(2M)^4 \epsilon_*^2}  \left[8 \alpha[\frac{16}{243} -\frac{40}{243}\log(3 \mu \sqrt{\epsilon_*}/4)] + \beta \frac{32}{81} - \gamma \frac{160}{243}\right]. \label{pth}
\end{eqnarray}

The renormalization mass $\mu$ is expected to be smaller than the Planck scale , while in the physically relevant regime,  $\epsilon_* \sim M^{-1}$. Thus, the logarithmic terms in Eqs. (\ref{rrr}---\ref{pth}) are of order $\log (\mu/\sqrt{M})$. For a solar mass black hole and taking $\mu = 1$, $\log (\mu/\sqrt{M}) \simeq -44$. Thus, for macroscopic black holes, the assumption that
$\log (\mu/\sqrt{M})  < - 10$ is very conservative. Given values of $\alpha$ and $\beta$ as in Eq. (\ref{alphabeta}), this assumption implies that the logarithmic term dominates and renders all expectation values $\bar{\rho}_*,  \bar{P}^r $ and $\bar{P}^{\theta}$ positive.

If we interpret the expectation values of the stress-energy tensor as thermodynamic densities and pressures, these correspond to an anisotropic fluid with different pressures in the radial and in the tangential direction.
However, these densities and pressures are positive. Therefore, even if they are included into the TOV equation (generalized for anisotropic fluid), they cannot lead to the  formation of a horizon. Hence, the semi-classical Einstein Eqs. (\ref{backreaction}) for  back-reaction {\em cannot} describe an equilibrium black hole. A different method is needed that will provide a resolution to the problems of the semi-classical Einstein equations that we listed earlier.

We believe that the best method for the treatment of quantum back-reaction for equilibrium gravitating systems  is to incorporate the quantum effects, including the trace anomaly, into the thermodynamic description of the system. This means that we should redefine the entropy functional in order to include contributions from the quantum effects associated to the trace anomaly. Then, we can construct an equation of state that takes these corrections into account and employ the classical Einstein's equations for this new equation of state. It is essential that this description is thermodynamically consistent; the
   constitutive equations of the system including back-reaction should correspond to maximum entropy solutions given boundary conditions similar to the ones employed in this paper.

\section{Conclusions.}

We showed that the horizon of an equilibrium black hole is surrounded by a thin shell where the Einstein equations break down. The existence of the shell follows from the requirement that classical GR coexists with the quantum effect of Hawking radiation in a consistent thermodynamic description.
 The shell has
  proper width $l \sim \sqrt{M}$, temperature $T_{sh} \sim 1/\sqrt{M}$, internal energy $E \sim \sqrt{M}$ and entropy $S_{sh} \sim M$. The proper width of the shell is much larger than the Planck length. Hence, the breakdown of the equations of GR is fundamentally not due to quantum gravity effects, but due to the quantum properties of matter (radiation).
The  shell's properties are independent of the box radius $R$. This  strongly suggests that these properties persist even when  the box is removed and  the system evolves slowly out of equilibrium, i.e., to   Schwarzschild black holes.



We emphasize the robustness of our conclusions.   We made no assumptions about the quantum characteristics of the underlying theory (unitarity, CTP symmetry, holography). We placed no restrictions on the origin of the new physics near the horizon. In deriving the properties of the shell, we used {\em only} thermodynamic principles and classical  GR. Nonetheless, the results are consistent with QFT in curved spacetime, in the sense that the regime in which quantum effects become significant is consistent with the shell properties derived by the maximum entropy principle.




\appendix
\section{  Analytic evaluation of the AH solutions}

We present an approximate analytic expression for the AH solutions that is valid in the regions $\alpha$ and $\beta$ of Fig. 1.

An AH solution is characterized by  $v_R << u_R$. In the region $I$,  $u$ increases with decreasing $r$ and $v$ decreases with decreasing $r$. Hence, the condition $v << u$  applies to all points in region $I$.
By continuity, the condition $v << u$ also applies in  a part of  region $II$.

In what follows, we denote derivative with respect to $\xi$ by a prime.

For $v << u$, Eq. (\ref{equv}) becomes
\begin{eqnarray}
u' = - u \hspace{2cm}
v' = \frac{2v (1-2u)}{1-u} \label{du12}
\end{eqnarray}
Hence,
\begin{eqnarray}
\frac{dv}{du} = - \frac{2v (1-2u)}{u(1-u)}, \label{dvdua}
\end{eqnarray}
The solution of Eq. (\ref{dvdua}) with the boundary condition $v(u_R) = v_R$,
\begin{eqnarray}
v = \frac{v_R u_R^2 (1 - u_R)^2}{u^2 (1-u)^2}. \label{vu1}
\end{eqnarray}
 Eq. (\ref{du12}) implies that $u(\xi) = u_R e^{-\xi}$. Substituting into Eq. (\ref{vu1}), we derive
\begin{eqnarray}
v(\xi) = \frac{v_R  (1 - u_R)^2 e^{2\xi}}{ (1 - u_R e^{-\xi})^2} \label{vx1}
\end{eqnarray}

Next, we study the AH solution in the regime where $1 - u(\xi) <<1$.   For sufficiently small $v_R$,
this condition applies to the whole of region $II$ and to a segment of  region $I$.

We set $u = 1 - \epsilon$. For $\epsilon << 1$, Eq. (\ref{equv}) is approximated by
\begin{eqnarray}
\epsilon' &=& 1 - 2v \\ \label{de1}
v' &=& -\frac{2v (1 +\frac{2}{3}v)}{\epsilon}. \label{de2}
\end{eqnarray}

Eqs. (\ref{de1}--\ref{de2}) imply that
\begin{eqnarray}
\frac{d \epsilon}{dv} = -\frac{\epsilon (1-2v)}{2v (1 +\frac{2}{3}v)} \label{veb}
\end{eqnarray}
Eq. (\ref{veb}) has solutions of the form
\begin{eqnarray}
\frac{v}{(v+\frac{3}{2})^4} = \frac{a}{\epsilon^2}, \label{vu1a}
\end{eqnarray}
for some constant $a$.

The maximum value of $u$ occurs for  $\xi = \xi_*$, such that $u'(\xi_*) = 0 $, or equivalently $v(\xi_*) = \frac{1}{2}$. The surface  $\xi = \xi_*$ is the approximate horizon $O_*$.
  We denote by $\epsilon_* = \epsilon(\xi_*)$ the blue-shift parameter on the approximate horizon.
    Eq. (\ref{vu1a}) implies that $a = \epsilon_*^2/32$. Then, Eq. (\ref{vu1a}) becomes
\begin{eqnarray}
\frac{32v}{(v+\frac{3}{2})^4} = \left(\frac{\epsilon_*}{\epsilon}\right)^2. \label{vu2}
\end{eqnarray}
Using Eqs. (\ref{vu2}) and (\ref{de2}), we obtain a differential equation for $v(\xi)$
\begin{eqnarray}
(v^{-1/2} + \frac{3}{2} v^{-3/2}) v' = - \frac{16 \sqrt{2}}{3 \epsilon_*}.
\end{eqnarray}
Integrating from some reference point $\xi = \xi_r$ with $v(\xi_r) = v_r$, we find

\begin{eqnarray}
2(\sqrt{v(\xi)} - \sqrt{v_r}) - 3 \left(\frac{1}{\sqrt{v(\xi)}} - \frac{1}{\sqrt{v_r}}\right) = \nonumber \\ - \frac{16 \sqrt{2}}{3 \epsilon_*} (\xi - \xi_r) \label{vx2}
\end{eqnarray}

Eqs. (\ref{vu1}) and (\ref{vu2}) have different, but not disjoint,  domains of validity.   For sufficiently small $v_R$,
both approximations are valid in a neighborhood of  the point  $P$ (see, Fig. 1).

Comparing Eqs. (\ref{vu1}) and (\ref{vu2}) near $P$, we find that
\begin{eqnarray}
\epsilon_* = \frac{16}{9} u_R (1 - u_R) \sqrt{2 v_R}, \label{e0b}
\end{eqnarray}
which relates the  parameter $\epsilon_*$  defined on  the approximate horizon $O_*$ to the boundary variables $u_R$ and $v_R$.

 Eq. (\ref{e0b}) implies the following relation between the boundary temperature $T_R$ and the parameter $\epsilon_*$.
 \begin{eqnarray}
T_{R} = \frac{3 \sqrt{\epsilon_*}}{4 \sqrt{2M} (8 \pi b)^{1/4}\sqrt{1 - u_R}}. \label{bt0b}
\end{eqnarray}

Using Eq. (\ref{vx2}) for a choice of the reference point $\xi = \xi_r$ lying in the domain of validity of Eq. (\ref{vx1}),
\begin{eqnarray}
\xi = \log u_R + \frac{3 \epsilon_*}{16 \sqrt{2}} \left( \frac{3}{\sqrt{v(\xi)}} - 2 \sqrt{v(\xi)} \right). \label{xv}
\end{eqnarray}
Setting $\xi = \xi_*$ in Eq. (\ref{xv}), we obtain
\begin{eqnarray}
\xi_* = \log u_R + \frac{3 \epsilon_*}{8}.
\end{eqnarray}

Using the radial coordinate $r = R e^{\xi}$, we identify the radial coordinate $r_*$ at the approximate horizon   to leading order in $\epsilon_*$
\begin{eqnarray}
r_* = 2M \left( 1 + \frac{3 \epsilon_*}{8} \right). \label{r*}
\end{eqnarray}
The corresponding value of the mass function $m_* = m(r_*)$ is
\begin{eqnarray}
m_* = \frac{1}{2} r_*( 1 - \epsilon_*) = M \left( 1 - \frac{5 \epsilon_*}{8} \right). \label{m*}
\end{eqnarray}

We evaluate the metric in the vicinity of $O_*$. To this end, we write the metric Eq. (\ref{ds2}) as
\begin{eqnarray}
ds^2 = - f(r)dt^2 + h(r) dr^2 + r^2 (d\theta^2 +\sin^2\theta d \phi^2),
\end{eqnarray}
in terms of functions $f(r)$ and $h(r)$. We express these functions as a Taylor series with respect to $(r - r_*)$ by computing their values and the values of their derivatives at $O_*$. To this end, we use Eqs. (\ref{equ}---\ref{equv}), for $u'$ and $v'$, which we differentiate successively in order to obtain equations for all derivatives of $v$ and $u$. Expressing $f$ and $h$ in terms of $u$ and $v$, we compute the derivatives of $f$ and $h$ at $r = r_*$. We list the derivatives that are necessary for the calculation of the trace anomaly.

\begin{eqnarray}
f_* &=& \frac{9}{16} \epsilon_*, \hspace{2.5cm} h_* = \epsilon_*^{-1},  \label{fh*1}
\\
 f_*' &=& \frac{3}{4} (2M)^{-1} , \hspace{1cm} h_*'= 0,
 \\
 f_*'' &=&  \frac{1}{2}(2M)^{-2} \epsilon_*^{-1}, \hspace{1cm}  h_*'' = -\frac{8}{3} \epsilon_*^{-1} (2M \epsilon_*)^{-2},
 \\
 f_*''' &=&  -\frac{2}{3}(2M)^{-3}\epsilon_*^{-2}, \hspace{0.6cm} h_*''' = \frac{80}{9} \epsilon_*^{-1} (2M \epsilon_*)^{-3},
 \\
 f_*'''' &=&  \frac{2}{3} (2M)^{-4} \epsilon_*^{-3} \hspace{1cm}h_*'''' = \frac{32}{3} \epsilon_*^{-1} (2M \epsilon_*)^{-4}. \label{fh*5}
\end{eqnarray}

 The lapse function $N = \sqrt{f}$ near $O_*$ is
 \begin{eqnarray}
N = \frac{3\sqrt{\epsilon_*}}{4} + \frac{1}{4Μ\sqrt{ \epsilon_*}} (r - r_*) + \ldots \label{lapse}
 \end{eqnarray}

The acceleration $a_i = \nabla_i \log N$ is purely radial, with
$a_r = \frac{1}{3M \sqrt{\epsilon_*}}$
Hence, the proper acceleration $a = \sqrt{a_{\mu} a^{\mu}}$ at $O_*$ is
\begin{eqnarray}
a =  \frac{1}{4M},
\end{eqnarray}
i.e., it equals the surface gravity of a black hole of mass $M$.

\section{ Evaluating the radiation entropy}

We evaluate the entropy of radiation in the regions I and II of an AH solution, Eq. (\ref{srad}).
For solutions to the Oppenheimer-Volkoff equation, the integrand in Eq.(\ref{srad}) is a total derivative, i.e.,
\begin{eqnarray}
\frac{r^{1/2} v^{3/4}}{\sqrt{1-u}} = \frac{d}{dr} \left( \frac{v + \frac{3}{2}u} {6v^{1/4}\sqrt{1-u}} r^{3/2}             \right).
\end{eqnarray}

Hence, $S_{rad} = S_1 - S_*$ where
\begin{eqnarray}
S_1 = \frac{2}{9} (4 \pi b)^{1/4} \frac{v_R + \frac{3}{2} u_R}{v_R^{1/4} \sqrt{1-u_R}} R^{3/2}
\end{eqnarray}
depends on field values at the boundary $r = R$, and
\begin{eqnarray}
S_* &=& \frac{4}{9} (8 \pi b)^{1/4} \frac{1 - \frac{3}{4}\epsilon_*}{\sqrt{\epsilon_*}} r_*^{3/2} \nonumber \\
&\simeq& \frac{4}{9} (8 \pi b)^{1/4} (2M)^{3/2} \left( \epsilon_*^{-1/2} - \frac{3}{16}\epsilon_*^{1/2}\right)
\end{eqnarray}
depends on the field values at $r = r_*$. Using Eq. (\ref{e0b}) to eliminate $v_R$ from $S_1$, we obtain
\begin{eqnarray}
S_{rad} = \frac{(8 \pi b)^{1/4}}{12}  (2M)^{3/2} \sqrt{\epsilon_*} (1 + \frac{9}{16 u_R^3 (1 - u_R)^2} \epsilon_*). \label{srad3}
\end{eqnarray}

Eq. (\ref{srad2}) applies in the regime
 $
K << 1$,
 where
\begin{eqnarray}
K = \frac{R^3}{4 \pi^2 M^4(1 - u_R)^2},
 \end{eqnarray}
so that the second term inside the parenthesis in the r.h.s. of Eq. (\ref{srad3}) is negligible for the value of $\epsilon_*$ that maximizes the total entropy, Eq. (\ref{estar}).


\end{document}